\documentclass{elsarticle}
\usepackage{amsmath}
\usepackage{graphicx}
\usepackage{hyperref}
\hypersetup{         
    pdftoolbar=true,        
    pdfmenubar=true,        
    pdffitwindow=false,     
    pdfstartview={FitH},   
    pdftitle={My title},    
    pdfauthor={Author},     
    pdfsubject={Subject},   
    pdfcreator={Creator},   
    pdfproducer={Producer}, 
    pdfkeywords={keywords}, 
    pdfnewwindow=true,      
    colorlinks=true,       
    linkcolor=blue,          
    citecolor=blue,        
    filecolor=magenta,     
    urlcolor=cyan           
}

\usepackage[section]{placeins}

\newcommand{\error}{\mu_\mathrm{e}}
\begin{document}

\title{Disentangling trust from cooperation: Evolution of trust as reduced monitoring in social dilemmas}

\author{Cedric Perret$^{1}$}
\author{The Anh Han$^{2}$}
\author{Elias Fern\'{a}ndez Domingos$^{3,4}$}
\author{Theodor Cimpeanu$^{5}$}
\author{Simon T. Powers$^{6}$}

\maketitle
	{\footnotesize
		\noindent
        $^{1}$ Department of Economics, University of Lausanne, Switzerland (\texttt{cedric.perret.research@gmail.com})\\
        $^{2}$ School Computing, Engineering and Digital Technologies, Teesside University, U.K. (\texttt{T.Han@tees.ac.uk})\\
         $^{3}$ Machine Learning Group, Universit\'e libre de Bruxelles, Belgium\\ 
        $^{4}$ AI Lab, Vrije Universiteit Brussel, Belgium (\texttt{elias.fernandez.domingos@vub.be})\\
        $^{5}$ Division of Biological and Environmental Sciences, University of Stirling, U.K. (\texttt{theodor.cimpeanu@stir.ac.uk})\\
        $^{6}$ Division of Computing Science and Mathematics, University of Stirling, U.K. (\texttt{s.t.powers@stir.ac.uk})\\
        \noindent  $^\star$ Corresponding author: Simon T. Powers (\texttt{s.t.powers@stir.ac.uk})
	}

\section*{Abstract}
\noindent It is commonly assumed that trust increases cooperation. However, game-theoretic models often fail to distinguish between cooperative actions and trust, making it difficult to independently measure trust and determine how its effects vary in different social dilemmas. To address this, we build on influential theories that equate trust with reduced monitoring of an agent's actions. We implement this as a heuristic that cognitively bounded agents can use in repeated games to avoid spending time and effort always monitoring their partner. Agents using this heuristic reduce monitoring of a partner's actions once a threshold level of cooperativeness has been observed -- providing a measurable and architecture-agnostic definition of trust. Using evolutionary game theory, we systematically analyse the success of strategies that use this trust heuristic across the entire space of two-player symmetric social dilemma games. We demonstrate that trust-as-reduced-monitoring facilitates cooperation in two different ways. First, when monitoring is costly, trust heuristics allow for higher levels of cooperation in social dilemmas where the temptation to defect is high. Second, when agents can make action errors, trust heuristics promote cooperation even in coordination problems. Our results disentangle trust from cooperation, and provide a behavioural measure of trust that applies across interaction types.
\vspace{3mm}

\noindent\textbf{Keywords:} Trust, Cooperation, Costly monitoring, Social dilemmas, Evolutionary game theory. 
\newpage

\section{Introduction}
Trust underlies cooperation in every human society \cite{fukuyama1995trust,botsman2017can}, and hence has been called ``the glue of healthy societies and the grease of economic productivity'' \cite{wef:2020:a}. This implies that the Artificial Intelligence (AI) agents of the future will also need the capability to build trust, to effectively cooperate \cite{perc_statistical_2017,capraro_mathematical_2021,battiston_higher-order_2025} with each other and with humans \cite{dafoe_cooperative_2021,hammond_multi-agent_2025}. However, this raises a fundamental question: what is the individual adaptive benefit of trusting another agent? By trusting another, an agent exposes itself to risk \cite{Luhmann:1979:a} -- the risk that the other agent might not act in an honest and reliable way.  

This question has been difficult to answer because there are many different verbal definitions of trust in both social science and artificial intelligence. These definitions can be categorised into: 1. behavioural definitions, which equate trust with more cooperative choices in a game, and 2. psychological definitions, which equate trust with psychological states such as viewing another individual as benevolent and acting with integrity \cite{lewicki_trust_2011}. Prior game-theoretic work on trust in economics \cite{ho2021trust} and multi-agent systems \cite{ramchurn_trust_2004} has typically followed 1., and effectively equated trust with high levels of cooperation. The risky investment decision modelled by the Trust Game \cite{Berg1995TrustHistory,kumar_evolution_2020,liu_evolutionary_2026}, which is effectively a one-sided Prisoner's Dilemma \cite{buskens_social_1998}, provides a paradigm example. The fundamental problem with this is that it conflates an outcome, cooperation, with one of its causes, trust \cite{james_jr_trust_2002,yamagishi_separating_2005,banu_how_2024}. As such, these definitions of trust do not allow us to delineate the role of trust in promoting cooperation. An alternative tradition in multi-agent systems has followed 2., and equipped agents with cognitive mechanisms to make trust decisions \cite{falcone_social_2001}. However, this has still not resulted in a general and mechanistic way to measure trust that applies in a wide range of settings. This is because it defines trust relative to an agent’s internal state, which may not be observable in an interpretable way, e.g. as is the case for current Large Language Model (LLM) agents, as well as for human agents. 

To overcome the problems with both 1. and 2., we need a measure of trust that is behavioural, so that it can be observed regardless of an agent’s architecture (artificial or human), but that does not conflate trust with cooperation. In this paper, we develop a behavioural measure derived from the adaptive function of trust behaviours, i.e. the individual benefits that trust brings to an agent. We do this by using evolutionary game theory (EGT) to formalise a key theory of trust -- that trusting means reducing monitoring of another agent's actions. This provides a population-level measure of trust -- the frequency of conditional strategies that reduce monitoring. This measure of trust is behavioural, since it only requires knowing whether an agent performs monitoring actions, but does not make reference to an agent's internal state. Consequently, it is agnostic to agent architecture.

Trust as reduced monitoring fits the intuitive folk psychology rule of ``If I trust you then I don't have to keep checking what you're doing''. This is in accordance with theories from social science; in particular, Luhmann argues that human agents use a trust heuristic to avoid having to calculate the likelihood of every possible outcome of an interaction \cite{Luhmann:1979:a,lewicki_trust_2011}. When trust is established, the adaptive benefit is that people save the time and effort of trying to exhaustively calculate whether or not they should cooperate and take the risk of being exploited. This has often been hinted at in the definitions of trust used in artificial intelligence \cite{loi_how_2023}. For example, in Taddeo's influential definition of e-trust, greater levels of trust entail less effort spent supervising the trusted agent \cite{taddeo_modelling_2010}. Similarly, in Castelfranchi and Falcone's notion of strict trust, trust is antagonistic to monitoring, where monitoring is defined as an action ``aimed at ascertaining whether another action has been successfully executed or if a given
state of the world has been realized or maintained...'' \cite[pg. 193]{castelfranchi_trust_2010}. Trust as a decision to reduce resources (time and effort) invested into monitoring has also been proposed as a pragmatic definition of trust between people and artificial intelligence \cite{ferrario_ai_2020,han_when_2021,ferrario_trust_2021,ferrario_being_2025,zahedi_game-theoretic_2025}. 

Previous work has modelled trust-as-reduced-monitoring using EGT in the context of a repeated Prisoner's Dilemma game \cite{han_when_2021}. However, this was restricted in scope to interactions that can be modelled as Prisoner's Dilemmas, which excludes many common situations where the temptation to defect is less, such as coordination problems. We go beyond this work by providing a game-theoretic formalisation of the trust-as-reduced-monitoring heuristic that applies to many types of interaction settings, while cleanly separating cooperation from trust. To do this, we build on the theory of repeated games.

In repeated games, agents interact with each other several times (rounds). This gives agents the possibility to condition their behaviour on how the other agent has acted previously, for example, by using reciprocal strategies such as ``cooperate this time if the other agent cooperated on the previous round''. These reciprocal strategies -- where agents can base their knowledge of how the other agent has acted previously on either their own first-hand knowledge (direct reciprocity) or information from third parties (indirect reciprocity) \cite{xia_reputation_2023}  -- underpin cooperation in human societies \cite{lehmann_four_2022}. But crucially, to be able to use reciprocal strategies, an agent needs to know whether the observed actions of another agent are actually cooperative or not. It is traditionally assumed that once an agent observes an action, it can instantly and with zero cost determine whether that observed action was cooperative. In other words, monitoring the actions of a co-player is costless. 

However, in reality, monitoring whether an observed action was actually cooperative carries an \emph{opportunity cost}, representing the time, effort and resources spent that could be allocated to other productive activities. For example, in e-commerce it may not be easy for a buyer to tell whether a seller sent them goods of the advertised quality, rather than say a fake, and so choosing to monitor this may take a substantial amount of their time. Although reputational information can sometimes reduce the need for this kind of direct monitoring (see, e.g., reputation mechanisms in Trust Games \cite{hu_adaptive_2021,xia_costly_2022}), reputational information itself can need verification for accuracy. For instance, determining whether reviews about a seller or service provider are genuine or fake can be difficult and time-consuming, and these costs may increase in the future given the ability of LLMs to automate the generation of plausible-sounding review text \cite{shukla_fighting_2024,powers_whats_2025}. Finally, monitoring the actions of AI agents can also be very costly, as this may involve simulating execution of their source code \cite{kovarik_game_2023}, the cost of which would depend on the power of the simulating agent and the auditing frameworks in place. Thus, in most real-world settings, conditional strategies such as Tit-for-Tat would need to pay a monitoring cost in order to know if their co-player cooperated on the previous round or not.

In classic game theory, some models have considered cases where agents can choose whether to pay a cost to reveal their co-players' actions, and have demonstrated folk theorem results about the existence of cooperative equilibria under particular conditions, such as communication \cite{matsushima_theory_1991} or the presence of a random signal \cite{miyagawa_folk_2008,hino_efficiency_2019}. However, these models have focussed on equilibrium outcomes rather than competition amongst different strategies employed by cognitively bounded agents. Consequently, they have not connected a monitoring cost with an adaptive trust heuristic. 

We address this gap by operationalising trust-as-reduced-monitoring into an agent's decision-making mechanism, through introducing a class of \emph{trust-based strategies} into repeated games. In contrast to traditional reciprocal strategies that implicitly monitor every round, such as Tit-for-Tat \cite{key:axelrod81}, trust-based strategies only occasionally monitor their partner's actions once a trust threshold has been reached. This trust threshold occurs after they have monitored their co-player for a number of rounds and have found their actions to be cooperative. Trust-based strategies thus reduce the opportunity cost of monitoring their partner's actions every round. 

Our results show the extent to which a trust-as-reduced-monitoring heuristic, as embodied by trust-based strategies, both evolves and promotes cooperation in a wide range of social interactions. We analyse the full space of symmetric two-player social dilemma games \cite{wang2015universal,Santos2006EvolutionaryPopulations}, and settings where agents can make unintentional mistakes in their actions. We find that trust-as-reduced-monitoring most strongly promotes cooperation in situations where there is a large temptation to defect. In contrast, this trust heuristic has the least effect on promoting cooperation in coordination interactions (Stag-Hunt games), although it still evolves in these cases. But in all cases, including coordination, we find that trust-as-reduced-monitoring is beneficial in promoting cooperation if there is a moderate probability of agents making unintentional errors in their actions. Overall, our results show that a wide range of repeated social dilemmas select for the evolution of a trust heuristic that avoids always monitoring another agent's actions, and that this promotes cooperation where 1. there is a strong temptation to defect, or 2. agents can make action errors.
    
\section{Models and methods}
We first describe the model and then present the methods used to analyse it. We follow the presentation in the \textit{Models and methods} section of \cite{han_when_2021}, but extend the model to cover different types of games, and introduce the possibility of agents making errors in their actions.

\subsection{Pairwise social dilemmas} 
We model the interactions between individuals as a pairwise social dilemma, defined by the following parameterised payoff matrix 
(for the row player), as in \cite{wang2015universal,Santos2006EvolutionaryPopulations}: 
\[
 \bordermatrix{~ & C & D\cr
                  C & R & S \cr
                  D & T & P  \cr
                 }.
\]
An agent who chooses the cooperate (C) action when their partner chooses the defect (D) action receives the sucker's payoff $S$, while the agent who defects gains the temptation payoff, $T$. Mutual cooperation (resp., defection) gives a reward $R$ (resp., punishment $P$) for both agents. This payoff matrix defines the stage game, i.e. the payoffs that result from a single round of interaction. 

Depending on the ordering of the four payoffs, different social dilemmas arise \cite{key:macy2002, Santos2006EvolutionaryPopulations}. We use the following scaling of the payoff matrix, where 
$R = 1$, $P = 0$, $0 \leq T \leq 2$, $-1 \leq S \leq 1$. 
Varying $S$ and $T$ captures all possible pairwise social dilemmas. Namely (see \cite{Santos2006EvolutionaryPopulations}): 
\begin{itemize}
    \item Prisoner's Dilemma (PD): $T > 1 > 0 > S $;
    \item Snowdrift (SD): $T > 1 > S > 0 $ and $T+S<2$; 
    \item Stag-Hunt (SH): $1 > T > 0 > S $. 
\end{itemize}

In a repeated game, agents have strategies. These are functions that map from the information that the agent has about the actions its partner took in previous rounds, to the action that the agent will take in the current round. In a population of $N$ individuals playing a repeated social dilemma as defined above, the fitness of an individual with a strategy \textbf{A} in a population with $k$ \textbf{A}s and $(N-k)$ \textbf{B}s is given by  
\begin{equation} 
\label{eq:PayoffA} 
\Pi_A(k) = \frac{1}{r(N-1)}\sum_{j=1}^{r}[(k-1)\pi_{A,A}(j) + (N-k)\pi_{A,B}(j)],
\end{equation} 
where $\pi_{A,A}(j)$ and $\pi_{A,B}(j)$ denote the payoff obtained by the A individual in a single round, $j$, when it interacts with an individual of strategy A (B, respectively), as defined by the payoff matrix above. The symbol $r$ denotes the total number of rounds in the game. Rather than considering a fixed number of rounds that agents know in advance, we instead make the standard assumption that the game continues for another round with probability $w$. This results in the expected number of $r =(1-w)^{-1}$ rounds per interaction \cite{key:Sigmund_selfishnes}. Throughout this paper, we compute all values of $\Pi$ analytically.

\subsection{Conditional strategies and the opportunity cost of monitoring a co-player's actions}

In repeated games, agents can condition their strategies on the previous actions of their co-players. Here, we introduce a monitoring cost, denoted by $\varepsilon$, which an agent must pay to know whether their partner made a cooperative action. The standard Tit-for-Tat (TFT) strategy (``Cooperate on the first round; in subsequent rounds, do the action that the co-player did in the previous round'') pays this cost every round. At the other extreme, fully unconditional strategies, i.e. always cooperate (AllC) or always defect (AllD), never pay this cost.

\subsubsection{Trust-based strategies}
Previous work introduced the concept of trust-based strategies for a repeated Prisoner's Dilemma game \cite{han_when_2021}. The TUC (Trust-based Cooperation) strategy begins by playing TFT and monitoring the action of its partner every round (paying the cost $\varepsilon$ to do so). However, it considers the difference between the number of cooperate versus defect moves its partner has been found to make. Once this exceeds a threshold $\theta$ (known as the trust threshold), then on each subsequent round it cooperates unconditionally with probability $1-p$ (and hence does not pay $\varepsilon$). On the other hand, with probability $p$ it monitors the partner to determine whether their last action was cooperate or defect. If the partner is found to have defected, it reverts to playing TFT for the remaining rounds (see also Figure~1 of \cite{han_when_2021}). 

We also consider a strategy TUD (trust-based defection), which is designed to exploit TUC. The TUD strategy begins by cooperating, but once the trust threshold $\theta$ is reached, switches to playing defect unconditionally (see also Figure~2 of \cite{han_when_2021})

The payoff matrix for the five strategies AllC, AllD, TFT, TUC and TUD can be written as follows:
\begin{equation}
\label{Eq:payofmatrix}
\bordermatrix{	
	~ 		& \textbf{AllC} & \textbf{AllD} &	\textbf{TFT} & 
	\textbf{TUC}	& \textbf{TUD}	 \cr
	\textbf{AllC} 	& 1 
					& S 
					& 1 
					& 1 
					& \frac{\theta + (r - \theta)S}{r} 
					\cr
	\textbf{AllD} 	& T 
					& 0 
					& \frac{T}{r} 
					& \frac{T}{r} 
					& \frac{T}{r} 
					 \cr
	\textbf{TFT} 	& 1 - \varepsilon 
					& \frac{S}{r} - \varepsilon 
					& 1 - \varepsilon 
					& 1 - \varepsilon 
					& \frac{\theta + S}{r} - \varepsilon
					\cr 
	\textbf{TUC}	& 1 - \frac{\theta\varepsilon}{r} - \frac{p(r- \theta) \varepsilon}{r} 
					& \frac{S}{r} - \varepsilon 
					& 1 - \frac{\theta \varepsilon}{r} - \frac{p(r- \theta) \varepsilon}{r} 
					& 1 - \frac{\theta \varepsilon}{r} - \frac{p(r- \theta) \varepsilon}{r} 
					& \Pi_{TUC,TUD} 
					\cr
	\textbf{TUD} 	& \frac{\theta + (r - \theta)T - 
						\theta\varepsilon}{r} 
					& \frac{S}{r}-\varepsilon 
					& \frac{\theta  + T - \theta 
						\varepsilon}{r} 
					& \Pi_{TUD,TUC}
					& \frac{\theta - 
						\theta\varepsilon}{r} 
					\cr
}.
\end{equation}

For ease of exposition, in the above payoff matrix the payoff for TUC interacting with TUD is written as $\Pi_{TUC,TUD}$, and the payoff for TUD against TUC is written as $\Pi_{TUD,TUC}$. These are defined as follows (as in \cite{han_when_2021}), 
\begin{align*}
     \Pi_{TUC,TUD}  
    &= \frac{\theta(1-\varepsilon)}{r} 
+ \frac{1}{rp} \left[ S\left( 1 - (1-p)^{\,r-\theta} \right) 
- \varepsilon \left( (1-p)^{\,r-\theta} + (r-\theta)p - 1 \right) \right], \\
\Pi_{TUD,TUC} &=
\frac{1}{r} \left[ \theta (1 - \varepsilon) + \frac{T \left( 1 - (1-p)^{r - \theta} \right)}{p} \right].
\end{align*}
These can be understood as follows. In the first $\theta$ rounds, before the trust threshold is reached, both TUC and TUD play C and keep monitoring their partners' actions. In each of these rounds they therefore both obtain a payoff $R - \varepsilon$. Once the trust threshold $\theta$ is reached, TUC will monitor the action of its partner only occasionally with probability $p$, while TUD will switch to defecting. Suppose that in the next round TUC monitors its partner's action and so finds them to be defecting. In this case, it obtains $S$ in the current round and $P-\varepsilon$ in the remaining rounds, since it reverts to TFT. Otherwise, i.e. if TUC does not monitor in that round (with probability $1-p$), the process above is iterated for the payoffs calculation.   

\subsubsection{Error rate}

\begin{figure}
\centering
\includegraphics[width=\linewidth]{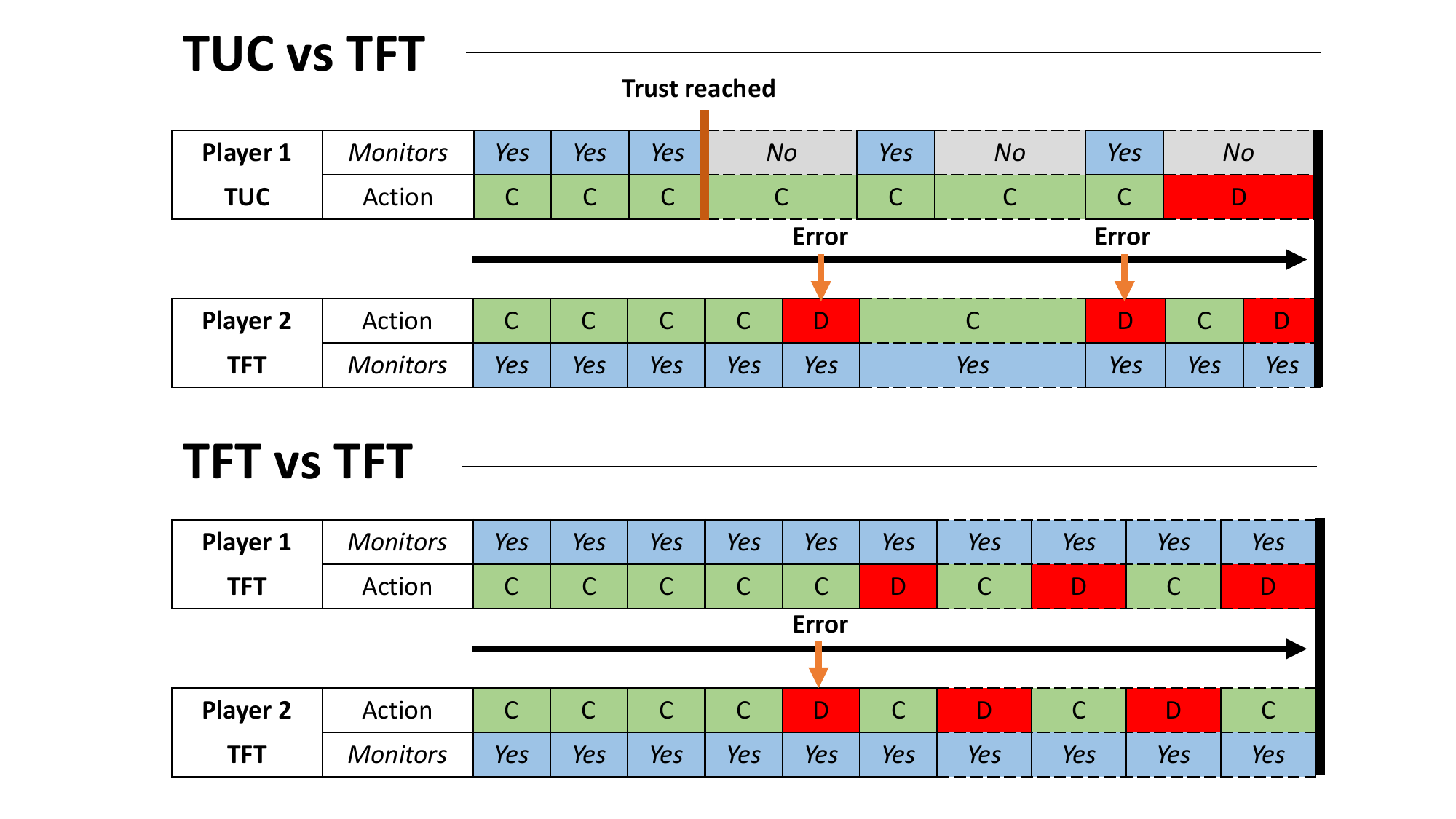}
\caption{\textbf{Trust-based Cooperation (TUC) can be more robust to errors than Tit-for-Tat (TFT).} The top panel shows a TUC agent playing against a TFT agent. Both agents start cooperating, and the trust threshold is reached by TUC. After a period of time, TFT makes an error and defects by mistake. However, because TUC has not monitored its co-player's action on this round, this error is overlooked and mutual cooperation resumes for a number of rounds. At a later stage, TFT defects by mistake again. This time, TUC does monitor, and reverts to playing TFT for the remaining rounds. By contrast, the lower panel shows that when TFT plays against itself, the first error always causes the breakdown of mutual cooperation.}
\label{fig:TUC_error}
\end{figure}

We also allow for the possibility that agents can make unintentional errors in their actions. We assume that with probability $\error$, an agent performs the action opposite to the one given by its strategy (i.e. defects instead of cooperates, or vice versa). Figure~\ref{fig:TUC_error} illustrates how both TUC and TFT respond to these errors. As the figure illustrates, TUC can be more resilient to these errors if they happen after the trust threshold has been reached, since they will be ignored if they occur on rounds where TUC does not monitor.

\subsection{Evolutionary dynamics in finite populations}
In this section, we follow the presentation of the method in Section~2.4 of \cite{han_when_2021}. To determine conditions under which trust-based strategies are advantageous, we use EGT \cite{MaynardSmith:1982:a,key:Sigmund_selfishnes}. Unlike classic game theory, EGT avoids the need to assume that agents are fully rational, but instead considers a population of agents that interact (play the repeated game), receive a total payoff (also called fitness), and can then change their strategy through payoff-biased social learning. This gives rise to evolutionary dynamics where more successful strategies, with higher average payoffs from the game, increase in frequency over time as a result of being copied by less successful agents. 

We model this social learning process using a pairwise comparison rule in which an agent $A$ (with fitness $f_A$), copies the strategy of another agent $B$ (with fitness $f_B$) with a probability given by the Fermi function \cite{Traulsen2006StochasticFixation}, $$P_{A \rightarrow B} = \left(1 + e^{-\beta(f_B-f_A)}\right)^{-1}.$$ The parameter $\beta$ gives the intensity of selection, i.e. how strongly an agent bases its decision to copy the strategy of another agent on the fitness difference between itself and the other agent. When $\beta=0$ the decision to copy the other agent's strategy is random, but as $\beta$ increases copying a higher fitness strategy becomes more deterministic. 

In this paper, we consider evolution in a finite population of $N$ agents. In finite populations, stochastic effects can have significant effects on evolutionary dynamics \cite{nowak2004emergence,rand2013evolution,zisis2015generosity,hauert2007via}. Because of stochasticity, lower-payoff strategies may sometimes spread through the population by chance, whereas higher-payoff strategies may die out. This stochastic framework is effective in accounting for empirical findings from human behavioural experiments \cite{rand2013evolution,zisis2015generosity}, making it an appropriate method to analyse our model.

Without behavioural exploration or mutations, the end states of evolution must be monomorphic, with only one strategy present. Whenever such a state is reached, it cannot be escaped via social learning. Therefore, we also assume that, with some mutation probability, an agent can freely explore its behavioural space and hence switch to a strategy not currently being used by other individuals in the population. If we take the limit of small mutation rates then the evolutionary dynamics can be described by a Markov chain, where each state represents a monomorphic population, and the transition probabilities between states are given by the fixation probability of a single mutant. This Markov Chain has a stationary distribution, which gives the average time the population spends in each one of these monomorphic end states. 
Notably, this approximation remains valid even beyond the strict limit of infinitely small mutation or exploration rates \cite{Traulsen2006StochasticFixation,hauert2007via,key:Hanetall_AAMAS2012}.

Let us suppose that there are at most two strategies present in the population, e.g. $k$ agents using strategy A ($0 \leq k \leq N$) and $(N-k)$ agents using strategy B. 
We denote by $\pi_{X,Y}$ the payoff an agent using strategy $X$ obtains in an interaction with another agent using strategy $Y$ (as defined by the payoff matrix (\ref{Eq:payofmatrix})).
The average payoff to an agent that uses strategy A (B, respectively) can be written 
\begin{equation} 
\label{eq:PayoffB}
\begin{split} 
\Pi_A(k) =\frac{(k-1)\pi_{A,A} + (N-k)\pi_{A,B}}{N-1},\\
\Pi_B(k) =\frac{k\pi_{B,A} + (N-k-1)\pi_{B,B}}{N-1}.
\end{split}
\end{equation} 
   
The probability of changing the number $k$ of agents using strategy A by $\pm \ 1$ in each time step as a result of social learning is given by \cite{Traulsen2006StochasticFixation} 
\begin{equation} 
T^{\pm}(k) = \frac{N-k}{N} \frac{k}{N} \left[1 + e^{\mp\beta[\Pi_A(k) - \Pi_B(k)]}\right]^{-1}.
\end{equation}
The fixation probability of a single strategy A mutant in a population of $(N-1)$ strategy B agents is \cite{Traulsen2006StochasticFixation,Karlin:book:1975,key:imhof2005}
\begin{equation} 
\label{eq:fixprob} 
\rho_{B,A} = \left(1 + \sum_{i = 1}^{N-1} \prod_{j = 1}^i \frac{T^-(j)}{T^+(j)}\right)^{-1}.
\end{equation} 
For a set $\{1,...,s\}$ of distinct strategies, these fixation probabilities determine the transition matrix of the Markov Chain: $M = \{T_{ij}\}_{i,j = 1}^s$, with $T_{ij, j \neq i} = \rho_{ji}/(s-1)$ and  $T_{ii} = 1 - \sum^{s}_{j = 1, j \neq i} T_{ij}$. The normalised eigenvector of the transpose of $M$ associated with the eigenvalue 1 gives the stationary distribution described above \cite{key:imhof2005}. This defines the relative time the population spends adopting each strategy.

\section{Results}
We begin our analysis by clarifying the role of reciprocal strategies in promoting cooperation across the range of pairwise social dilemmas defined by the Prisoner's Dilemma, the Snowdrift, and the Stag-Hunt games. We then determine how reciprocal strategies are affected by the realistic assumption of a cost of monitoring a partner's actions, $\varepsilon>0$, and how the effect of the monitoring cost differs between games. From this, we proceed to analyse the evolutionary advantage of trust-based (reduced monitoring) strategies, and their effectiveness in promoting cooperation across the different social dilemma games. We initially assume, unless otherwise specified, that $\error=0$, i.e. agents do not make errors in their actions, before relaxing this assumption in Section~\ref{secErrors}.

\subsection{When does reciprocity promote cooperation in different social dilemmas?}

\begin{figure}
\centering
\includegraphics[width=\linewidth]{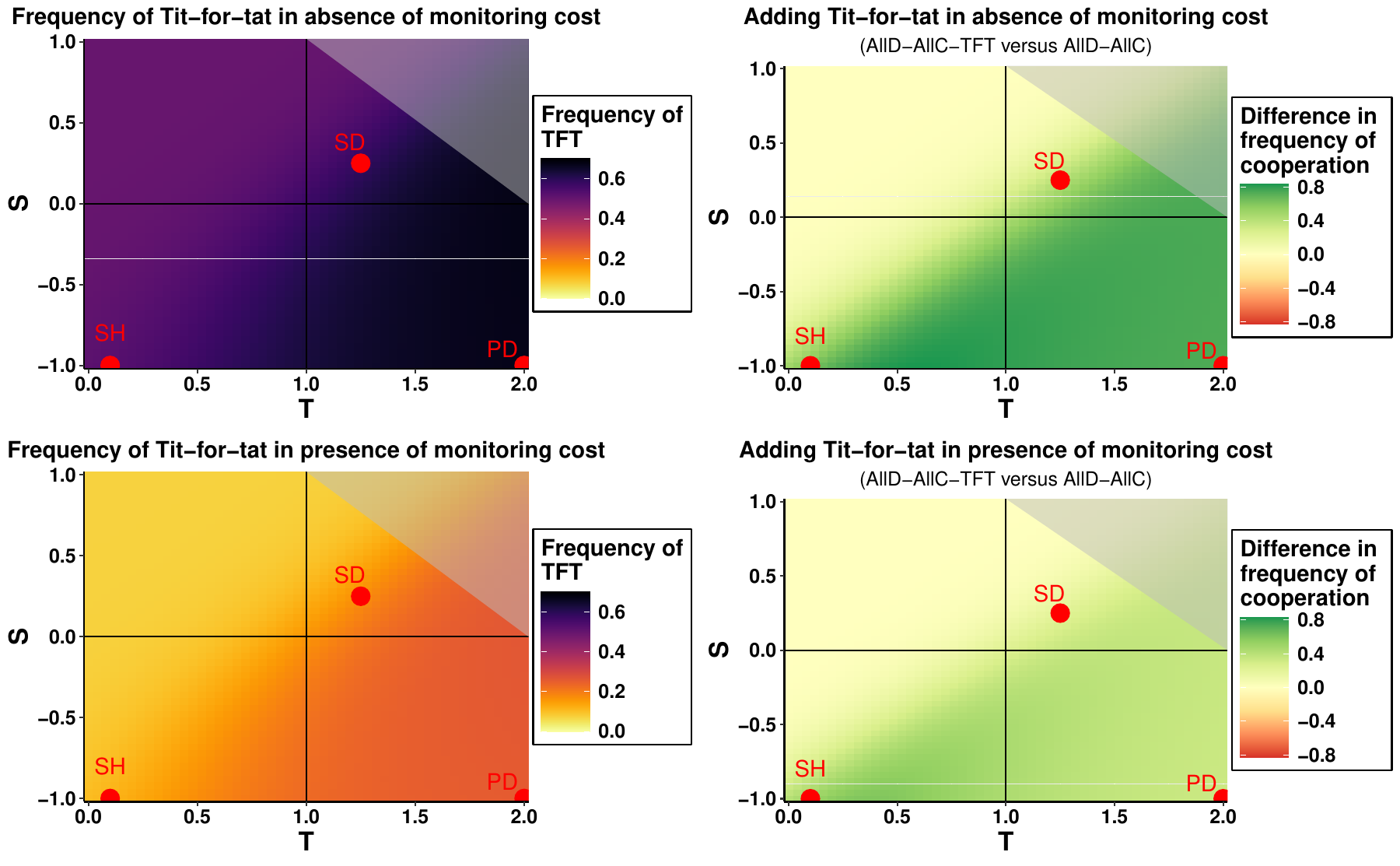}
\caption{\textbf{Frequency of the Tit-for-Tat (TFT) strategy (left), and the impact on the frequency of cooperation of adding TFT to the pool of available strategies AllC and AllD (right)}. The results are shown as a function of the temptation to defect $T$ and sucker's payoff $S$, and in the absence ($\varepsilon=0$, top panels) or presence ($\varepsilon=0.25$, bottom panels) of the cost of monitoring a partner's actions. Prisoner's Dilemma (PD) games occur in the lower right quadrant of $S$-$T$ space, Snowdrift (SD) games occur in the top right quadrant and Stag-Hunt (SH) games occur in the lower left quadrant. The red dots show the particular game instances that we use for later analysis. Parameters: a game with $r = 50$, and evolutionary dynamics with $\beta = 0.1$ and $N = 100$.}
\label{fig:adding_TFT}
\end{figure}

\subsubsection{The role of reciprocity across social dilemma games}  We first examine how Tit-for-Tat (TFT) performs across the space of Prisoner's Dilemma (PD), Snowdrift (SD), and Stag-Hunt (SH) games, and compute how much introducing TFT into a strategy pool that initially consists only of Always Cooperate (AllC) and Always Defect (AllD) promotes cooperation. We start by considering the case where there is no cost of monitoring a partner's actions. Our results, presented in the top part of Figure~\ref{fig:adding_TFT}, show that 
TFT is a frequent strategy across all social dilemma games, and is more frequent when the temptation to defect, $T$, is high. TFT is at the highest frequency in PD games, and less frequent in SD and in SH games. The top right panel in Figure~\ref{fig:adding_TFT} indicates that adding TFT into the pool of available strategies also increases the frequency of cooperation across all three social dilemma games.

\subsubsection{Tit-for-Tat and the cost of monitoring a partner's actions} Conducting the same analysis, but this time in the presence of a monitoring cost ($\varepsilon=0.25$), our results (bottom panels of Figure~\ref{fig:adding_TFT}) show that including a monitoring cost drastically reduces the frequency of TFT (compare the top left and bottom left panels). The drop in frequency is largest in SH situations, where TFT is close to absent for many SH games in $S$-$T$ space. By contrast, the drop in frequency of TFT is smallest for PD games. A monitoring cost also reduces the effect of the presence of TFT on increasing cooperation across all games. This effect is less pronounced in SH games, even though the frequency of TFT is lowest in this part of $S$-$T$ space compared to the other games. Thus, even a low frequency of TFT promotes cooperation in the risky coordination situations represented by the SH game.

\subsubsection{Cooperation results from Tit-for-Tat in Prisoner's Dilemma games, and from Always Cooperate and Tit-for-Tat in Stag-Hunt and Snowdrift games} To understand these findings, we look more closely at differences in the role played by reciprocity in these different games. Because cooperation results from either TFT or AllC individuals cooperating with one another, TFT strategies can promote cooperation in two ways. First, TFT can directly promote cooperation by simply being a frequent strategy. Second, it can indirectly promote cooperation by catalysing the evolution of the unconditional AllC strategy.

\begin{figure}
\centering
\includegraphics[width=\linewidth]{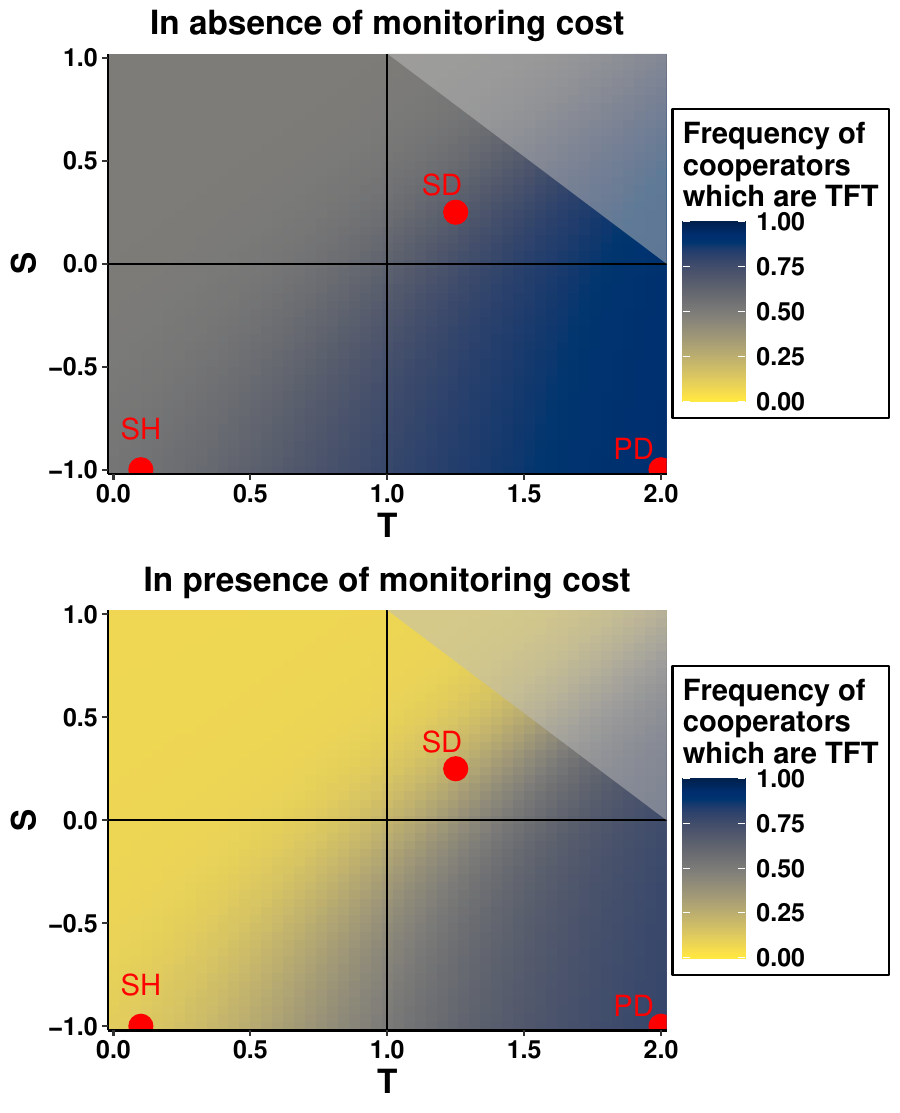}
\caption{\textbf{Frequency of cooperators using the Tit-for-Tat strategy}, in the absence of the cost of monitoring a partner's actions (top), and with the cost $\varepsilon=0.25$ (bottom). Parameters: a game with $r = 50$, and evolutionary dynamics where $\beta = 0.1$ and $N = 100$.}
\label{fig:adding_TFT_supp}
\end{figure}

Figure~\ref{fig:adding_TFT_supp} examines which strategies foster cooperation in different games, and demonstrates that in the PD, cooperation primarily arises from the high frequency of TFT. In contrast, in the SH and SD games, cooperation is due to a mix of AllC and TFT strategies. When considering costs of monitoring the partner's actions, this distinction becomes even more pronounced, with nearly all cooperation in the SH game resulting from the higher frequency of AllC. This finding explains why a monitoring cost has a strong impact in the PD, where cooperation results directly from TFT individuals cooperating with each other. But there is less impact in the SH, where cooperation results from AllC players that are immune to these costs since their behaviour is unconditional on their partner's actions. Now that we have established how TFT fares across different social dilemmas, and how a cost of monitoring a partner's actions affects this, we conduct the same analysis for trust-based strategies.

\subsection{Trust-as-reduced-monitoring across games}
We now investigate, as a function of the monitoring cost, (i) the frequency of the Trust-based Cooperation strategy TUC when all five strategies are possible, and (ii) the impact of adding trust-based strategies (TUC and TUD) to the pool of available strategies on the frequency of cooperation.

\begin{figure}
\centering
\includegraphics[width=\linewidth]{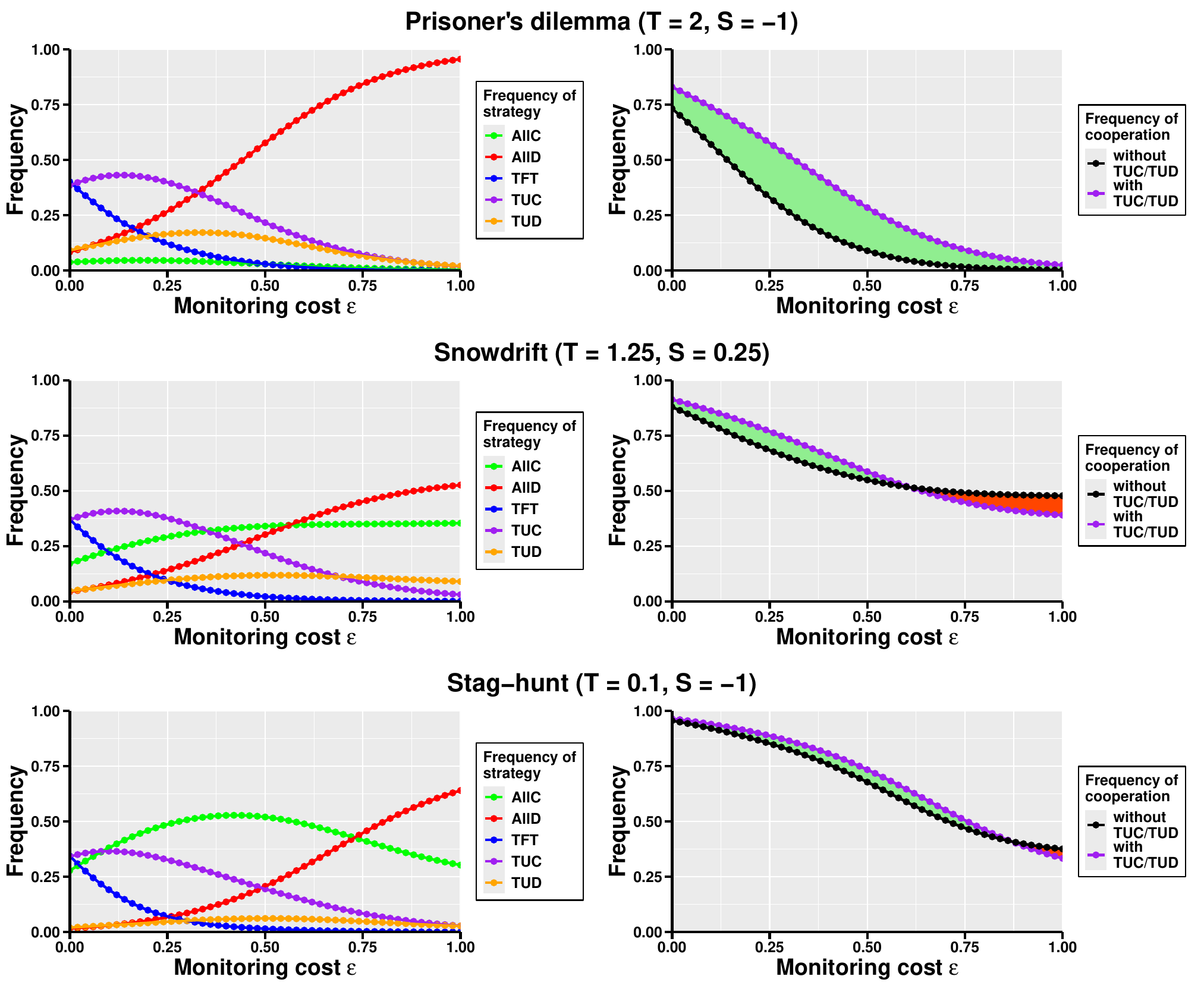}
\caption{\textbf{Left: Frequency of strategies as a function of the monitoring cost, $\varepsilon$. Right: Frequency of cooperation in the absence or presence of trust-based strategies TUC and TUD, as a function of the monitoring cost, $\varepsilon$}. The difference in frequency of cooperation when trust-based strategies are included is shaded in green where positive and red where negative. Results are presented for one example of each of the Prisoner's Dilemma, Snowdrift and Stag-Hunt games. Parameters: trust-based strategies with $\theta = 3$ and $p = 0.25$, games with $r = 50$, and evolutionary dynamics where $\beta = 0.1$ and $N = 100$.}
\label{fig:epsilon}
\end{figure}

\begin{figure}
\centering
\includegraphics[width=\linewidth]{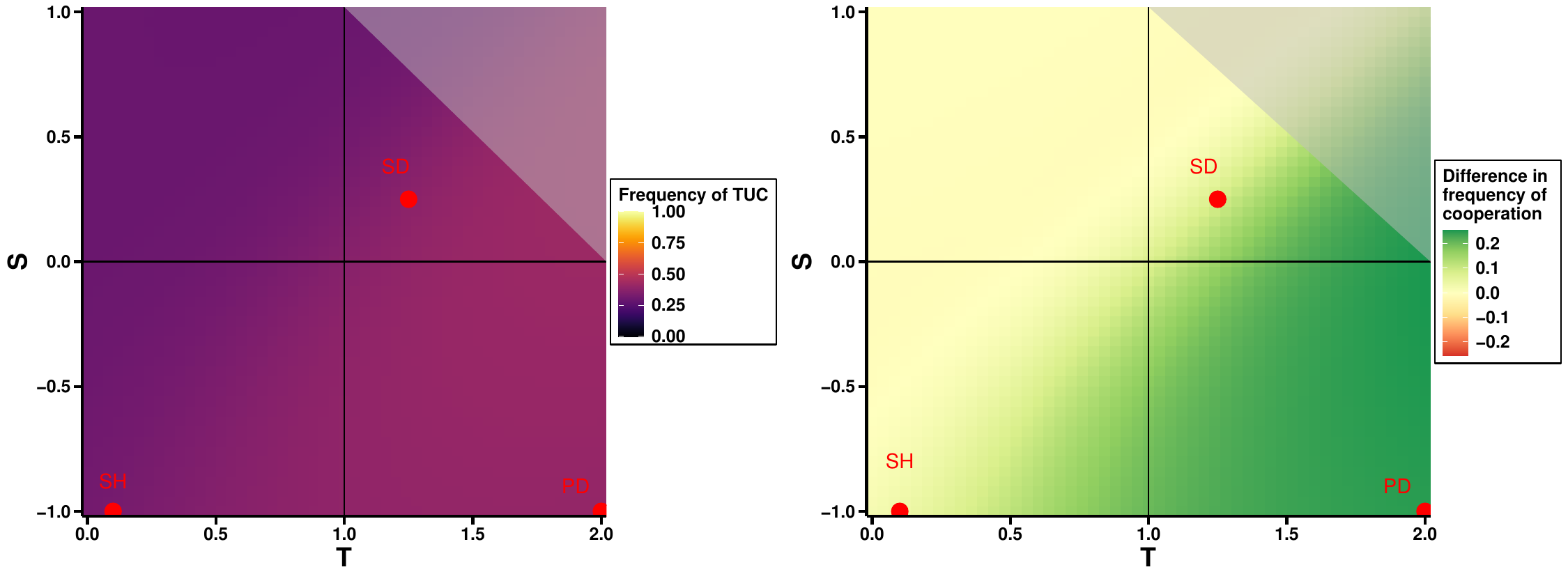}
\caption{\textbf{Frequency of the TUC strategy (left) and the impact of considering trust strategies on the frequency of cooperation (right), as a function of the temptation to defect $T$ and the sucker's payoff $S$.} The difference in frequency of cooperation is calculated between a system with all five strategies and a system with only AllC, AllD and TFT. The red dots correspond to the values of $S$ and $T$ used in Figure~\ref{fig:epsilon}. Parameters: trust-based strategies with $\theta = 3$ and $p = 0.25$, games with $\varepsilon = 0.25$ and $r = 50$, and evolutionary dynamics with $\beta = 0.1$ and $N = 100$.}
\label{fig:freq_coop_T_S}
\end{figure}

\subsubsection{Baseline analysis: Trust-based strategies are successful and promote cooperation in the repeated Prisoner's Dilemma} In the top part of Figure~\ref{fig:epsilon}, which focuses on the PD, our analysis replicates previous findings \cite{han_when_2021}. First, increasing the monitoring cost diminishes the frequency of the TFT strategy (top left panel), since TFT must pay the monitoring cost every round. Second, a higher monitoring cost favours TUC compared to TFT, with TUC being the most successful strategy overall for intermediate monitoring costs in the PD. Third, the inclusion of trust-based strategies (TUC and TUD) in a system that previously included only AllC, AllD and TFT leads to an overall increase in the frequency of cooperation (top right panel). When the monitoring cost is high, the AllD strategy becomes the most prevalent, but (i) TUC remains among the most common across a wide range of monitoring costs, and (ii) the inclusion of trust-based strategies always increases the frequency of cooperation for the range of monitoring costs studied.

\subsubsection{Trust-based Cooperation (TUC) is a frequent strategy across different games} Our setup now allows us to examine if these findings generalise across different types of games, and thus different types of social interaction. Our results in Figure~\ref{fig:epsilon} and the left panel of Figure~\ref{fig:freq_coop_T_S} demonstrate that Trust-based Cooperation (TUC) consistently remains a frequent strategy across the three different games, and for a broad range of monitoring costs. In other words, trust-as-reduced-monitoring is expected to be a widespread heuristic across various types of interaction, not just in PD situations.
A notable difference between the games is that trust-based strategies are highly prevalent in the PD, while AllC is almost absent. In this situation, cooperation with reduced monitoring after trust is established is a successful strategy, but unconditional cooperation is not. In the SD and SH games, both TUC and AllC strategies are frequent, indicating that a mix of careful trusters and unconditional cooperators could be observed in these more relaxed social dilemmas.

\subsubsection{The availability of trust-based strategies increases the frequency of cooperation when the temptation to defect is high} Our results reveal that the impact of incorporating trust-based strategies on cooperation differs across the various games. Specifically, while the availability of the Trust-based Cooperation strategy is always beneficial in the PD, this strategy has a limited effect on the frequency of cooperation in SD games and in SH games. As shown in Figure~\ref{fig:freq_coop_T_S}, trust-based strategies generally increase cooperation when the temptation to defect (T) is high. However, trust-based strategies do not seem to increase cooperation in SH coordination games, even though TUC is present at a significant frequency. This is because in coordination games, we have shown earlier that reciprocal strategies are only a springboard for individuals that always cooperate to become established. Once in place, these unconditional strategies can enjoy a high payoff without ever paying a monitoring cost.

\subsection{Trust-as-reduced-monitoring when individuals make errors}
\label{secErrors}
\begin{figure}[!h]
\centering
\includegraphics[width=\linewidth]{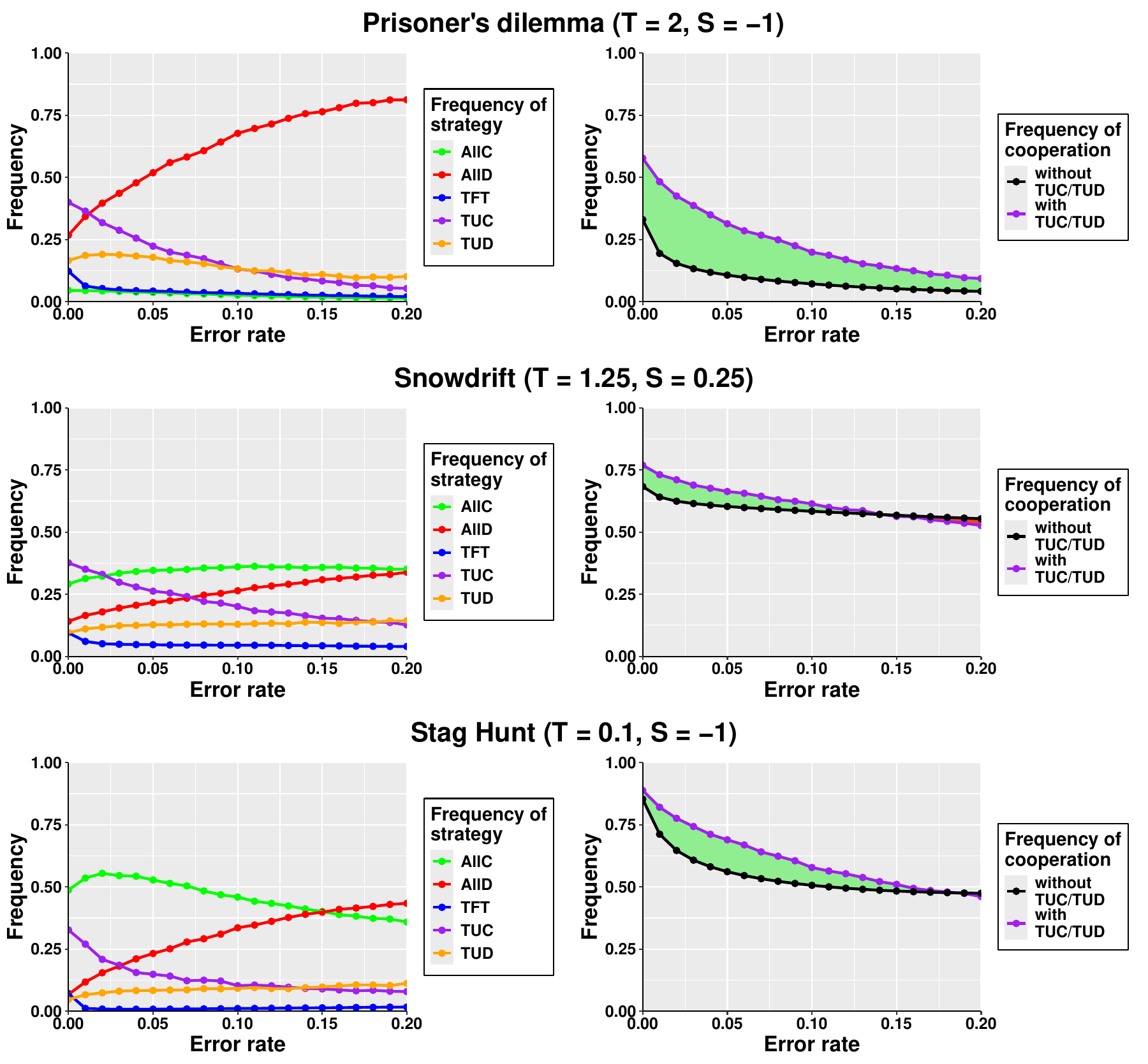}
\caption{\textbf{Frequency of strategies (left) and frequency of cooperation in the absence or presence of trust-based strategies TUC and TUD (right), as a function of the error rate, $\error$}. For clarity, the difference in frequency of cooperation is shaded in green when positive and red when negative. Parameters: trust-based strategies with $\theta = 3$ and $p = 0.25$, games with $\varepsilon = 0.25$ and $r = 50$, and evolutionary dynamics with $\beta = 0.1$ and $N = 100$.}
\label{fig:freq_error}
\end{figure}

Finally, we consider cases where agents can make unintentional errors in their actions. Figure~\ref{fig:freq_error} shows the frequency of the different strategies and the frequency of cooperation as a function of the error rate, $\error$. Our results show that an increasing error rate decreases the frequency of TUC across games, in favour of the strategy AllD. However, our results also show that the frequency of TUC remains considerable even for relatively large error rates such as 5\%. 

The right-hand side of Figure~\ref{fig:freq_error} shows that, in all games, error rates of up to 10\% do not lessen the beneficial effect of trust on the frequency of cooperation, and can even increase it. For instance, TUC increases cooperation the most in PD games when the error rate is around 2\%-3\%. Importantly, trust-based strategies improve cooperation when the error rate is greater than 0 in SH games, in contrast to the results where agents do not make errors. A trust-as-reduced-monitoring heuristic thus helps agents achieve cooperation in coordination dilemmas where errors can occur.

\begin{figure}[!h]
\centering
\includegraphics[width=\linewidth]{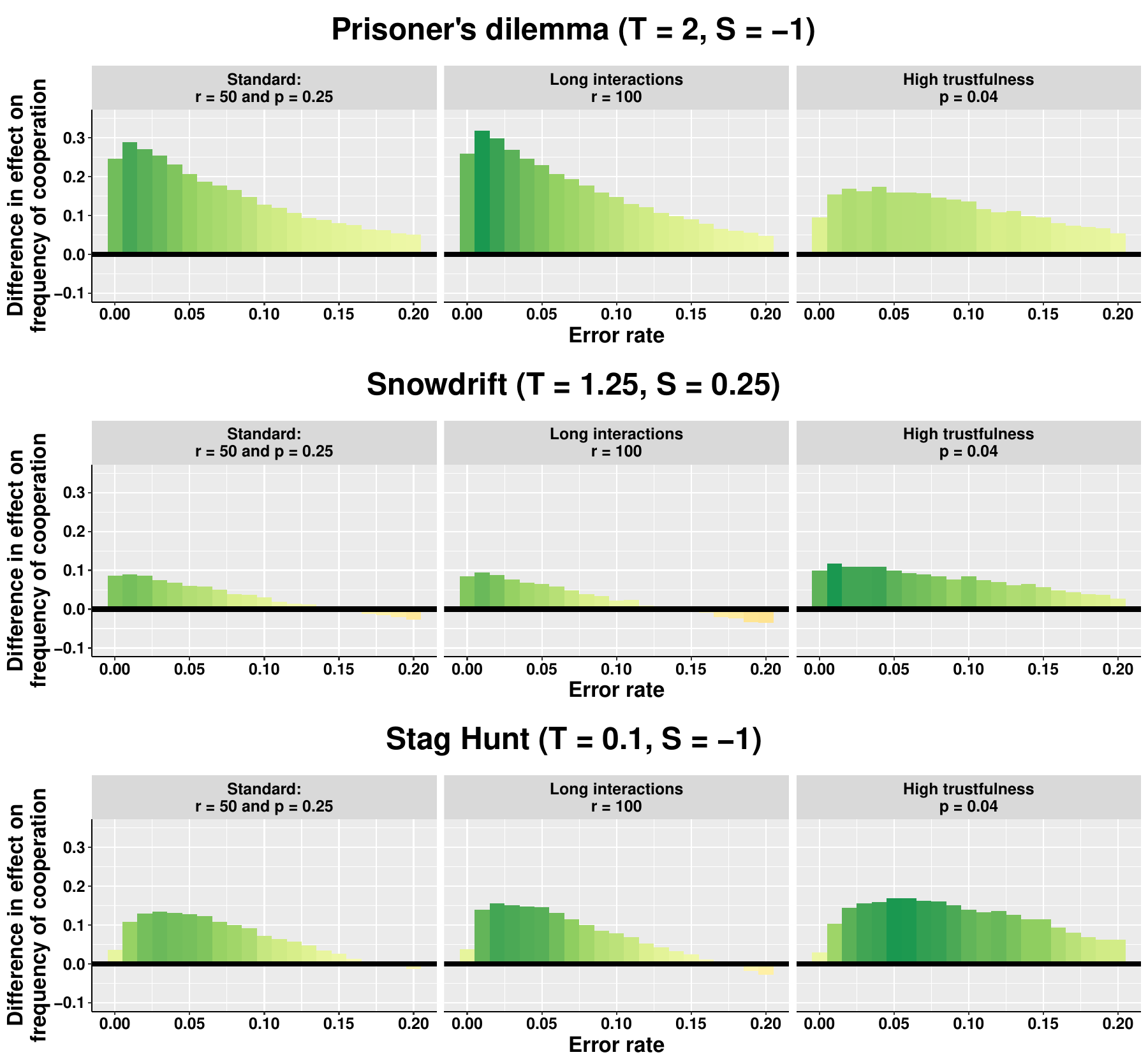}
\caption{\textbf{Difference in the frequency of cooperation when trust-based strategies TUC and TUD are included, as a function of the error rate, $\error$, for three different social dilemmas and for long interactions or high trustfulness}. Parameters: trust-based strategies with $\theta = 3$, games with $\varepsilon = 0.25$, and evolutionary dynamics with $\beta = 0.1$ and $N = 100$.}
\label{fig:error_games_p_r}
\end{figure}

Trust-based strategies are thus selected for in the presence of errors, and can be even more beneficial for cooperation than in situations where agents never make errors. Figure~\ref{fig:error_games_p_r} shows that these conclusions hold for games repeated for a larger number of rounds ($r=100$), and where trust-based strategies monitor less frequently once the trust threshold has been reached (high trustfulness, $p=0.04$). 

The effect of high trustfulness on cooperation varies with the type of game. Specifically, lowering $p$ from 0.25 to 0.04 increases cooperation in SD and SH games, but it decreases cooperation in the PD (right-hand panel of Figure~\ref{fig:error_games_p_r}). Thus, the optimum level of trustfulness for promoting cooperation depends on the type of social dilemma, with more relaxed dilemmas favouring greater trustfulness.

\section{Conclusion and implications}

Our results can be summarised by three points. First, we have shown that trust, as a heuristic to avoid the cost of always monitoring a partner's actions, is a successful strategy that can evolve in a wide range of interactions. Importantly, this happens even in the presence of exploitative strategies (TUD) tailored to take advantage of this trust.

Second, by disentangling behavioural measures of trust from behavioural measures of cooperation, we have been able to rigorously examine the effect that trust has on the overall level of cooperation in a population. Our results show that the presence of the Trust-based Cooperation strategy (TUC) increases the overall level of cooperation when the temptation to defect is high, but not in coordination dilemmas such as Stag-Hunt games. We have shown that this is because reciprocity in coordination games is useful to kickstart cooperation, but is not necessary once cooperation is established. This suggests that interventions aiming to promote the success of TUC, such as reducing high monitoring costs, would have less effect on cooperation in risky coordination games compared to stricter social dilemmas. 

Third, we have shown that when there is the possibility of agents making errors in their actions, TUC remains frequent and can even be more beneficial for cooperation compared to cases where agents never make errors. Notably, TUC does not suffer as much from errors as TFT, as TUC agents ignore errors from cooperative partners on the rounds that they do not monitor. By contrast, TFT is not trustful and monitors its partner's actions every round, which means that a single mistake causes TFT to defect in retaliation. This can cause the partner to defect in retaliation, leading to a loss of cooperation over many rounds resulting from one mistake \cite{nowak1992tit}. Trust can prevent this spiral into defection.

Our model is based on the assumption that it can be costly to know whether the action of another agent was actually cooperative -- even when you observe that action \cite{han_when_2021}. Everyday examples of this are common. For example, there is a time cost of monitoring if the correct amount has been debited from your bank account, or if goods that you have ordered online are genuine rather than fake. When interacting with artificial intelligence (AI), these costs can become even greater. For example, are personalised music recommendations from a streaming app influenced by paid promotions \cite{pelly2025mood}? Has an LLM agent completed a task in a way that meets the preferences I prompted it with, or were its actions not in my interests due to biases in its training data \cite{wu_unveiling_2024}? 

Our operationalisation of trust -- as choosing not to monitor another agent's actions -- provides an objective behavioural measure that applies across a broad spectrum of social interactions. Using this, trust can be measured in experimental settings by how often someone chooses (or not) to pay a cost to monitor another agent's actions. This could be carried out in standard behavioural economics settings \cite{camerer2003behavioral}, where a participant is given the option to pay to find out the action of its partner in each round of a repeated social dilemma game. Some experimental studies have already considered costly monitoring for the particular case of the Trust Game \cite{goeschl_trust_2017,ma_social_2020}. In this game, one player is the investor and the other player is the trustee. The investor begins with an endowment amount of money. They then choose a proportion (between 0 and 1 inclusive) of this to send to the trustee. The amount sent is multiplied by a factor $> 1$. The trustee then chooses the proportion of this to return to the investor. When the game is repeated, most experiments assume that the investor is given complete information about the previous actions of the trustee at no cost, i.e. that monitoring is costless. However, a study has shown that when a cost is introduced to obtain this information, people are sensitive to this cost and monitor less \cite{goeschl_trust_2017}. Moreover, the frequency of monitoring was lower in later rounds, in a pattern akin to the TUC strategy. Future work should extend these experiments beyond the trust game and into the larger space of social dilemma games that we have modelled.  

Crucially, this trust measure can also be applied in more realistic experimental settings. One important application of this is experiments to measure the factors that affect people's trust in AI. Although there have been many empirical studies in computer science that have aimed to measure factors affecting trust in various applications of AI, these studies have lacked a unified theoretical framework \cite{leichtmann_crisis_2022,leichtmann_effects_2023}. This has led to conflicting results and a lack of predictive models, with ``Trustworthiness mechanisms and measures [...] being advanced in AI regulations and standards that may not actually increase trust'' \cite{knowles_acm_2024}. The trust measure we have introduced here can help to resolve this impasse, by providing an objective measure of trust that is grounded in (evolutionary) game theory. For example, incentivised experiments could be designed where a user can take advice from an AI agent in a realistic setting, such as an e-commerce chatbot recommending products or an LLM answering questions about a text, and can pay a cost to verify (monitor) whether that advice is correct. This can then be connected with EGT models by varying factors such as how frequently the AI gives correct information (i.e. its strategy to cooperate or defect), or whether the AI has a high reputation from other users.

The costs of monitoring also lead to problems for the alignment and control of AI, that is, ensuring that the goals and behaviours of AI systems are aligned with human values and intentions. Research in AI safety has stressed the risk of scheming -- AI systems that deceptively appear to be aligned with human goals during training, but then pursue misaligned goals after deployment once they are no longer closely monitored \cite{meinke_frontier_2025}. This type of scheming corresponds to the TUD strategy in our model -- cooperate until an auditor or developer relaxes oversight, and then defect. The monitoring cost maps to the cost of auditing the AI system. Extensions of our model, with the payoff matrix calibrated to the parameters used in red-team blue-team AI control experiments \cite{greenblatt_ai_2024}, could therefore provide insights into auditing regimes and other institutional incentives that minimise vulnerability to scheming by LLMs and other future AI systems.

Most generally, our model demonstrates the evolutionary success of a trust heuristic in competition with strategies that do not use trust (AllC, AllD and TFT), and in competition with strategies designed to build and then exploit trust (TUD). Trust evolves even though trust-based strategies are costly to use compared to unconditional strategies (since they pay the monitoring cost before trust is established and periodically thereafter), and are risky compared to standard reciprocal strategies (since they could potentially be exploited once they stop monitoring every round). These results suggest that a trust heuristic could have evolved in humans by natural selection. Indeed, there is evidence that human trust behaviour is to some extent genetically determined \cite{riedl_biology_2012}, and hence is an evolved trait. Our results suggest that repeated social dilemma games, as played by humans throughout (pre)history \cite{powers_cooperation_2021}, could provide selection pressure for the evolution of trust.

There are several limits to our model which could be addressed in future work. First, we have not modelled the evolution of the parameters of trust-based strategies that control when they monitor their co-player's actions, $\theta$ and $p$. Rather, we have assumed that these are given exogenous values. Future work could allow these to be evolving traits, allowing evolutionary optimal values to be found for different game parameters.

Second, we assumed that lost trust cannot be gained again. In the real world, trust relationships can be rebuilt if interactions become cooperative again. This could potentially allow for greater long-term cooperation than is seen in our results, but could also open the door for more exploitative TUD style strategies. Exploitative strategies that take advantage of trust rebuilding would not necessarily be successful, though, because they  need to cooperate for a long period in order to rebuild trust before being able to exploit again. Moreover, this dynamic could drive the evolution of the parameters of Trust-based Cooperation, for example requiring a longer period of cooperation before trust is built (greater $\theta$), and a lower trustfulness (higher $p$). An important next step is therefore to consider the effect of more complex deceiving strategies on the form of trust that emerges. 

Third, we have modelled a well-mixed population of agents that are all similar in their cognitive capabilities. It would be important to examine the effects of spatial structure \cite{szabo2007evolutionary}, which can facilitate the evolution of cooperation in repeated games such as the Prisoner's Dilemma \cite{key:axelrod81}. Moreover, agents might differ in how easily they can monitor another agent's actions, and in the complexity of strategies that they can use. For example, when considering interactions between humans and AI agents, the model could be extended to allow AI agents to monitor actions at lower costs, and to use more complex strategies such as modelling and predicting human-partner trust behaviours. This would represent AI agents' greater access to information and processing power, respectively. This could be achieved by modelling a separate population of AI agents that compete with and copy strategies from each other, but earn payoffs from playing with a population of human agents.

Finally, the game-theoretic formalisation of trust as a cognitive shortcut to avoid the cost of continuous monitoring could apply beyond interactions that are repeated between the same pair of agents. These include $n$-player social dilemmas such as repeated public goods games. Trust-based strategies could also apply with indirect reciprocity, in cases where it is costly to get information from third parties about a partner's previous actions.   

In conclusion, we have developed a behavioural measure of trust that applies across the spectrum of symmetric two-player social dilemma games, and that distinguishes observed trust from observed cooperation. Because this behavioural measure does not depend on access to an agent's internal state, it can apply to both human and AI agents. Using the methods of evolutionary game theory, we have shown that Trust-based Cooperation, which monitors its partners' actions only occasionally after a trust-threshold is reached, is a successful strategy in a wide range of social interactions. We have also shown that the presence of trust increases the overall level of cooperation in a population, especially when agents can make unintentional errors in their actions. This adds formal support for the idea that high-trust societies lead to more cooperative and friction-free economies \cite{fukuyama1995trust}.  

\section*{Declaration of generative AI and AI-assisted technologies\\ in the manuscript preparation process}

During the preparation of this work the authors used AI-based tools (ChatGPT with Deep Research and Elicit) to assist with the identification of potentially relevant literature. All cited sources were independently read, evaluated and integrated into the manuscript by the authors. The authors take full responsibility for the content of the publication.

\section*{Acknowledgements}  T.A.H. is supported by EPSRC (grant EP/Y00857X/1).

\bibliographystyle{elsarticle-num}

\bibliography{Trust_nourl.bib}

\end{document}